# Excel as a Turing-complete Functional Programming Environment

Peter Bartholomew
MDAO Technologies Ltd
peter.bartholomew1@btinternet.com

**ABSTRACT**

*Since the calculation engine of Excel was the subject of a major upgrade to accommodate Dynamic Arrays in 2018 there has been a series of seismic changes to the art of building spreadsheet solutions. This paper will show the ad-hoc end user practices of traditional spreadsheets can be replaced by radically different approaches that have far more in common with formal programming. It is too early to guess the extent to which the new functionality will be adopted by the business and engineering communities and the impact that may have upon risk. Nevertheless, some trends are emerging from pioneering work within the Excel community which we will discuss here.*

## 1    INTRODUCTION

Until recently, spreadsheets have very much been the territory of the non-programmer. The majority of spreadsheets in use today do not even contain formulas. In fact, if one is to accept a remark made by David Benaim at a London Excel meetup, the practice of merging cells to assist the presentation of data is more widely encountered than the use of formulas.

When it comes to formulas, the current state of the art is still much as described by [Hellman, 2001] in his EuSpRIG paper, in which he blames most of the recognised deficiencies, such as the profusion of errors, auditing difficulties, lack of uniform development methodologies, and barriers to easy comprehension of the underlying business models, upon the standard spreadsheet user-interaction paradigm of the 'cell-matrix' approach. His solution was to develop an interface program that would accept formulas written in natural language and then generate the necessary spreadsheet formulas without further user intervention.

In this paper we will consider how equivalent development strategies are now possible working directly with built-in Excel formulas. The Excel developments that will be described start with the introduction of dynamic arrays, which were described by the present author at EuSpRIG [Bartholomew, 2019], then will proceed to address the impact of the LET function, as found in other functional programming languages, which allows the introduction of locally-scoped named variables to Excel. This was followed by Lambda which permits the development of user-defined functions without any need to resort to VBA or other Add-Ins. From the recording of the online POPL 2021 Conference presentation that featured Simon Peyton Jones, Andy Gordon, Jack Williams of Microsoft Research, it appears that the idea of turning the humble spreadsheet into a full-fledged programming language had been around for the best part of two decades before any implementation was undertaken, [Peyton Jones, 2021]. The claim of Turing completeness will not be demonstrated here but the implication is that it is now possible to program significant software algorithms just by using Excel spreadsheet formulas.

Finally, the paper will look at limitations of the current Excel implementation both in terms of functionality and the testing and verification that is essential to the management of risk.

## 2 THE EVOLVING SPREADSHEET LANDSCAPE

### 2.1 The traditional spreadsheet experience

Much of what we consider to be 'normal' spreadsheet practice dates all the way back to 1980s and ideas for the electronic spreadsheet developed within VisiCalc by Dan Bricklin and Bob Frankston. The idea of cells within a grid was inherited from manual paper-based spreadsheets but the A1 town plan notation for referencing cells was adopted simply to avoid the 'tedium' of the user having to define their terms using named variables. The aim was to allow users to work with a layout that was to be used for the final presentation of results which echoed the development of WYSIWYG word processing interfaces that had emerged only a little time earlier.

The early decisions very much saw the spreadsheet as an environment in which non-programmers could manipulate and present numerical data. That mindset largely persists up to this day. John Raffensperger [Raffensperger, 2001] argued that the idea of "the spreadsheet as a computer program" needs to be replaced and instead it should be viewed as a document that is intended to be read in order to communicate business logic and content. Much of his argument relating to layout are sound, especially in the prevailing context of the single-cell calculations based upon the concept of direct cell referencing.

Where I would disagree with Raffensperger in that "the spreadsheet as a computer program" is not a metaphor; it is the reality. To be in denial does not make the spreadsheet any less a program; it simply increases the likelihood of it being an extremely poorly constructed program. Despite the fact that arrays have been available from the time Excel was ported to Windows, years of experience answering questions posted to on-line forums would suggest that they are viewed as methods of last resort rather than a simplifying assumption that captures the significance of the calculation. Relatively few users have experience of array calculations, possibly because they are not discoverable; no one would stumble across the strategy of committing formulas by pressing Enter, whilst Control and Shift are held down, by accident! The traditional way of performing what should be an array calculation requires the introduction of the concept of relative references, so that a separate formula may be created for each element of the array by using copy/fill.

Again, defined names existed within past versions of Excel, but they are used relatively rarely and some standards like FAST [FAST, 2015] deprecate their use. Without defined names, formulas tend to read more like encrypted cyphers than natural language, so there is pressure to keep them short and ensure that they reference nearby cells. To find the significance of a reference, one must follow the precedence and then hope to find nearby annotation. In his 2016 paper, [Bartholomew, 2016] showed that it is possible to build an entire spreadsheet model using only CSE array formulas and how that, in turn, enabled the use of defined names, since one no longer needs to reference individual cells. As was pointed out during discussion, using techniques that were so far from normal practice would do little to address the problem of spreadsheet error because only a few developers would be in a position to deploy them.

Overall, the assumptions and techniques of the traditional spreadsheet have become so familiar that few would even recognise how much of an outlier spreadsheets are in the world of computing. There is no expectation that variables should be defined; formula references are specified solely by 'point and click' and provide no clue as to their business significance; the size of any array depends solely on how far the developer happened to drag the formula when setting it up. With modern Excel, functional programming concepts have entered the delightfully chaotic world of the spreadsheet.

### 2.2 Dynamic Array formulas

In September 2018 Microsoft announced the impending release of dynamic arrays within Excel. Prior to that, the value property of a single cell was limited to a single value, be it a String, a Double, a Boolean or an Error. At the Florida Ignite meeting Joe McDaid of Microsoft described how the newest releases of Excel could associate an array object with a single cell. Adjacent empty cells, known as a spill Range, would be used to display the result. The size of the spill range depends solely on the formula references, rather than user action, resulting in one element of risk being eliminated as a result.

An interesting fact that emerged at the time is that the calculation engine of Excel always was capable of calculation based upon the use of 2D arrays. It was just on the grid that a function that expected a scalar parameter but was given a vector, *e.g.*

```
= MOD(A1:A10, 3)
```

would first reduce the range reference to a single cell by implicit intersection and then return a single value. Had the same formula been specified within a defined name it would have been evaluated as an array formula. Similarly, if the range were converted to an array constant

```
= MOD({1;2;3;4;5;6;7;8;9;10}, 3)
```

the evaluation would have been correct. All that Ctrl/Shift/Enter did was remove the implicit intersection step. In modern Excel the default behavior is reversed and, to force Excel to perform the implicit intersection trick, one would need to use the '**@**' operator

```
= MOD(@$A$1:$A$10,3)
```

At the same time as these changes were introduced, several new array functions, including **SORT** and **FILTER**, were released. In legacy versions of EXCEL such operations required manual operations using ribbon commands but now the functions could be built into the automatic calculation.

Many spreadsheet users are not attracted to array formulas because they see anything other than scalar arithmetic as undesirably complicated. I would contend that the reverse is true; array formulas create amazingly simple solutions that structure the solution and replace a multitude of single cell formulas with a single readable formula. A hopeful sign is that Microsoft has reported a substantial upturn in the use of array formulas since the release of dynamic arrays. I now see posts that request dynamic range solutions rather than fill-down.

### 2.3 The LET function

Whereas, for me, the dynamic array formula had been long anticipated, the introduction of the **LET** function came as a complete surprise. In functional programming environments, **LET** introduces the scope in which local variables may be defined. The implementation in Excel is managed by having **2n+1** parameters in pairs, with odd numbered parameters being the variable names and even numbered parameters the formula that is to be evaluated.

It appears that programmers with greater experience of traditional Excel methods are likely to write the resulting formula as concisely as possible,

```
= LET(x,C5:C15=C3,y,FILTER(D5:D15,x),IF(y<>"",y,"-")),
```

whilst those who have embraced the new methods more completely may be more inclined to use a vertical layout

```
= LET(
      criterion,  account=required,
      selected,   FILTER(completionDate, criterion),
```

```
            IF(selected<>"", selected, "-")
        ),
```
where Alt/Enter (linefeed) is used to emphasise the code nature of the formula as a sequence of statements. Something particularly worthy of note is that the names in a **LET** statement are variables. That is, they are evaluated when first encountered and the value may then be used multiple times without recalculation. That is to be contrasted with the defined name which refers to a _formula_ that is re-evaluated every time it is encountered.

### 2.4   The LAMBDA function and recursion

The next step of introducing the **LAMBDA** function has far more profound implications but, apparently, it was reasonably easy to implement **LAMBDA** once **LET** was up and going. Rather than enabling the evaluation of variables within the formula, **LAMBDA** associates variable names with the terms of a list of function arguments and then passes the values by name into the body of the formula. To illustrate this, one might consider a simple exponential growth calculation throughout a period of time

```
    = ExponentialGrowthλ(10000, 5%, 12)
```

where the function is defined with the initial amount, growth per period and number of periods as parameters. The function itself is given by

```
    ExponentialGrowthλ
    = LAMBDA(initial, rate, nPeriods,
        LET(
                periods, SEQUENCE(1 + nPeriods, , 0),
                initial * (1 + rate) ^ periods
        )
    )
```

The formula serves to demonstrate the passing of parameters and, also, the manner in which the naming of Lambda functions can be used to express the purpose of the formula, whilst concealing the details of its calculation. Used well, this allows the developer considerable freedom to refactor the solution without the risk of generating errors as a result.

A more interesting problem arises when the growth factor changes from period to period, for example, as it would for a variable rate mortgage. One way of addressing that problem is through the use of recursion.

### 2.5   Recursion

Apparently, it is the use of recursion that has made modern Excel Turing complete; that is, capable of simulating the results of any other computer program. Whilst that may be true, it does not mean that recursion is easy to use, especially so given the skill profile of the typical spreadsheet user.

The worksheet formula

```
    = Recurλ(initialAmount, vRate)
```

uses a recursive Lambda function

```
Recurλ
= LAMBDA(opening, vRate, [p],
    LET(
        np,      COUNT(vRate),
        p',      IF(ISOMITTED(p), 1, p),
        closing, Growthλ(opening, vRate, p'),
        balance, IF(p' < np,
                Recurλ(closing, vRate, p'+ 1),
                closing),
        VSTACK(opening, balance)
    )
)
```

A key element of any recursive formula is the conditional statement that allows the function to call itself a finite number of times before exiting and running the remaining statements of the function at each level in the calling stack.

The other Lambda function found within the formula simply looks up the escalation factor for the specific period and applies it to the opening amount

```
Growthλ
= LAMBDA(rate, vRate, p,
    LET(
        rate,    INDEX(vRate, p),
        closing, opening * (1 + rate),
        closing
    )
)
```

For a while the creation of complex recursive formulae became something of an art form, but now they are now largely history, having been replaced by the much simpler-to-use Lambda helper functions.

### 2.6  LAMBDA helper functions

Like **LAMBDA** itself, the idea of helper functions is standard within functional programming languages. The functions are **MAP, BYROW, BYCOL, SCAN, REDUCE, MAKEARRAY**, each of which partitions an array parameter and feeds each part to a Lambda function as a parameter value, before reassembling the results back into an output array.

These helper functions provided a solution to several problems that had caused significant challenges to the use of dynamic arrays. An example of the first might be a dynamic array showing the projected value of several investments over future years. The task of summing the 2D array by row to get a column of annual total returns is trivial using relative referencing but was not so easy using dynamic array methods.

|  | Return | | |
|---|---|---|---|
|  | InvestmentA | InvestmentB | Yearly |
|  | $ 500 | $ 480 | $ 980 |
|  | $ 525 | $ 499 | $ 1,024 |
|  | $ 551 | $ 519 | $ 1,070 |
|  | $ 579 | $ 540 | $ 1,119 |
|  | $ 608 | $ 562 | $ 1,169 |
|  | $ 638 | $ 584 | $ 1,222 |
|  | $ 3,401 | $ 3,184 | $ 6,585 |

Figure 1. Projected investment return and calculated totals

This is easily done by a traditional fill-down formula

    = SUM($F10:$G10)

but, whilst the table is a dynamic array, a totals column calculated this way is not dynamic. There are dynamic array formulae that use **MMULT**, but the best solution is to use **BYROW**

    = BYROW(return#, LAMBDA(x, SUM(x)))

which feeds each row of investment returns into the **LAMBDA** function in turn. It may also be noted that the expression highlighted in red could be converted to a named Lambda function **Sumλ** (say), so enabling the calculation to be expressed more concisely as

    = BYROW(return#, Sumλ).

A more severe problem with dynamic array calculation arose with the task of creating running totals by accumulating flow variables. This is frequently referred to as a 'corkscrew' calculation. The problem is that any attempt to reference a previous term in a dynamic array returns a circular reference error.

| Date | February | March | April | May | June | July |
|---|---|---|---|---|---|---|
| Revenue | 105,000 | 110,250 | 115,763 | 121,551 | 127,628 | 134,010 |
| COGS | 135,000 | 125,000 | 115,000 | 105,000 | 95,000 | 85,000 |
| Cash balance | -30,000 | -44,750 | -43,988 | -27,437 | 5,191 | 54,201 |

Figure 2. Accumulating cash flows to give a cash balance

**SCAN** runs along a timeline and by defining and passing the values referenced by the second parameter to the Lambda function, **Addλ,** one at a time, where it is combined with the result from the previous step to give the new result. The formula used to accumulate the net cash flow is

    = SCAN(0, Revenue-COGS, Addλ)

where **Addλ** is defined by

    = LAMBDA(x, y, x + y)

At each step, **SCAN** will provide the previous result as **x** and the current flow value as **y**, which the function **Addλ** then sums.

The Lambda helper functions perform calculations that would otherwise require recursion but with a far more straightforward syntax. There are still shortcomings that prevent multiple accumulations to be performed simultaneously but, even there, workarounds involving **REDUCE** are possible.

## 2.7 Array shaping functions

The latest batch of functions are the array shaping functions. The traditional single-cell, manual approaches offer complete flexibility when it comes to the layout selected for the presentation of results, but array formulas are far more constrained. Two of the functions are used in the example showing forecast investment returns in the previous sub-section.

Figure 1, above, shows annual returns for two investments, but also shows a total value for each investment. At first sight one might expect this to generate spill errors as the number of years increases.

```
= LET(
    escalatedValue,  DROP(value#, -1) * escalation,
    investmentTotal, BYCOL(escalatedValue, Sumλ),
    blankRow,        {"",""},
    VSTACK(
        blankRow,
        escalatedValue,
        blankRow,
        investmentTotal
    )
)
```

The functions to note are **DROP** which removes the terminal values because only period opening values are used in the calculation of percentage growth. The overall return from each investment is calculated using the **BYCOL** function. Finally, the presentation is built using **VSTACK** with blank rows used to align and space the table of results.

These can do far more than profile output within the presentation layer. They can have a key role in the calculation process. For example, if you had a row of historical data by quarter, you could examine it for seasonality by wrapping the data by year and averaging over the years.

```
= LET(
    salesArray, WRAPROWS(DROP(sales#,,1), 4),
    BYCOL(salesArray / SUM(salesArray), Sumλ)
)
```

There are many other things that could be said about Lambda functions, including the fact that they are first class objects within Excel and can be passed as parameters to other Lambda functions. Unfortunately, providing a definitive treatment of Lambda lies beyond the scope of this paper.

What follows instead is a treatment of the types of Lambda usage that has emerged over the past two years and the needs they are intended to meet.

## 4 BUILDING SOLUTIONS USING LAMBDA FUNCTIONS

### 4.1 Categories of Lambda function

The initial focus when Lambda was introduced within Excel was on the opportunity it offered for the creation of UDFs without leaving the formula environment. Since then, a number of markedly different categories of Lambda have emerged. These include

- o Simple uses that turn built-in functions into Lambda functions for use with helper functions. These would include the **Sumλ** and **Addλ** functions used in the previous section. Another possibility might be to modify a built-in function, *e.g.* to return the sequence number of a record relative to a table, based upon the **ROW** function

    ```
    = Rowλ(record, table)
    ```

- o Functions that are problem-specific and employed to describe the process and intent of a particular solution. An example might be a function that distributes several recurring payments along a timeline

    ```
    LAMBDA(start, occurrences, periodicity, amount, counter)
      = LET(
          outflow, SEQUENCE(occurrences, , start, periodicity),
          BYCOL(IF(counter=outflow, amount), SUMλ)
        )
    ```

    where the names are based on the specific terms used within the workbook.

- o Lambda functions that extend the library of in-built functions of Excel by using published Lambdas. Examples of these are given in the next section, but can include complicated mathematical calculations such as optimisation.

- o Recently Craig Hatmaker [Hatmaker, 2023] has suggested that Lambdas could be used to generate major software components as the building blocks of a given class of related applications. The payoff would be both in terms of speed when building a new solution and the confidence provided by using pre-tested components.

Having described the plethora of change that Excel has been subjected to over the past few years, it is probably worth pausing to study some applications, since they bear little resemblance to anything that has preceded them.

### 4.2 Overhead crane motion using 4th order Runga-Kutta method

Although, in the literature, one more often sees examples of financial models than engineering models, over 85% of manufacturing OEMs and their tier 1 supply chain companies use Excel for calculations related to their products. The modelling in this example was intended to demonstrate the ability of Excel to perform highly mathematical calculations, given that claims are made as to its Turing completeness.

The problem considers the motion of a freight container under the action of control forces applied by the operator to the crane cart. The system of equations to be solved are

$$\dot{y} = v$$
$$\dot{\vartheta} = q$$
$$\dot{v} = \epsilon\vartheta + u$$
$$\dot{q} = -\vartheta - u$$

where $y$ and $v$ are the linear position and velocity of the cart, whilst $\vartheta$ and $q$ are angular measures. The control input is denoted by $u$.

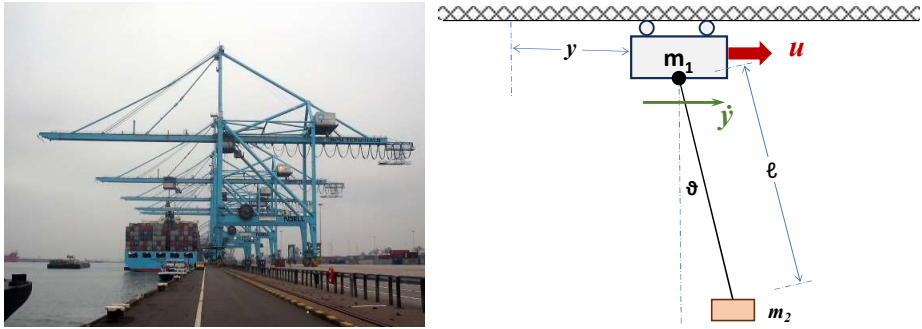

**Figure 3. Key parameters used to model crane and container dynamics**

The worksheet formula used for this is

    = SCANV(X₀, t, RK4Stepλ(Dλ))

where

    **RK4Stepλ**

    = LAMBDA(D, LAMBDA(xᵣ, tᵣ,
        LET(
            δx₁, δt * D(xᵣ, tᵣ),
            δx₂, δt * D(xᵣ + δx₁ / 2, tᵣ + δt / 2),
            δx₃, δt * D(xᵣ + δx₂ / 2, tᵣ + δt / 2),
            δx₄, δt * D(xᵣ + δx₃, tᵣ + δt),
            xᵣ₊₁, xᵣ + (δx₁ + 2 * δx₂ + 2 * δx₃ + δx₄) / 6,
            xᵣ₊₁
        )
    ))

It is likely that the normal spreadsheet user will not recognise the formula as having anything to do with Excel. The first step is to recognise the syntax of the **LET** function, with the first argument being a variable name and the next the formula is to be evaluated. The next point is that the names themselves look unfamiliar to an English speaker, though maybe less so to a Greek, but the Unicode characters can be accessed using Insert/Symbol and they form valid Excel Names. The names are constructed to look like the mathematical variables they implement, but one could equally use names like 'variables3rdTrial' if that makes one happier.

A further point to note is that **SCANV** is a Lambda function, written by the author, to generalise **SCAN** to output an array of arrays. **RK4Stepλ** is the implementation of the 4[th]-order Runga-Kutta algorithm shown above, but it is itself passed as a parameter to **SCANV** and accepts a Lambda function **Dλ** as a parameter. **Dλ** implements the dynamic equations, above, and are specific to the problem of the overhead crane.

The purpose of this section is to demonstrate that significant programming exercises may be conducted within the worksheet formula environment; it is not necessary to resort to VBA or JavaScript, and the worksheet formula environment is fast. Before leaving the section, it may be worth looking at the results obtained.

The loading sequence specified had been 2sec accelerating the cart, followed by 2 sec decelerating at half the intensity to allow the container to swing forward to catch up with the cart. The loading was then reversed, with 2 sec acceleration at half intensity before being brought to rest with a 2 sec deceleration at full intensity.

From the figure, below, it may be seen that this control loading sequence leads to substantial residual oscillations, making it difficult to set the container down at the end of the movement. The system was then optimised using Solver to improve the predicted behaviour.

It turned out that one parameter was sufficient to reduce the residual energy of the system to zero, so the durations of the load application was set, and the optimisation carried out with respect to the load intensity of the 2$^{nd}$ and 3$^{rd}$ control inputs.

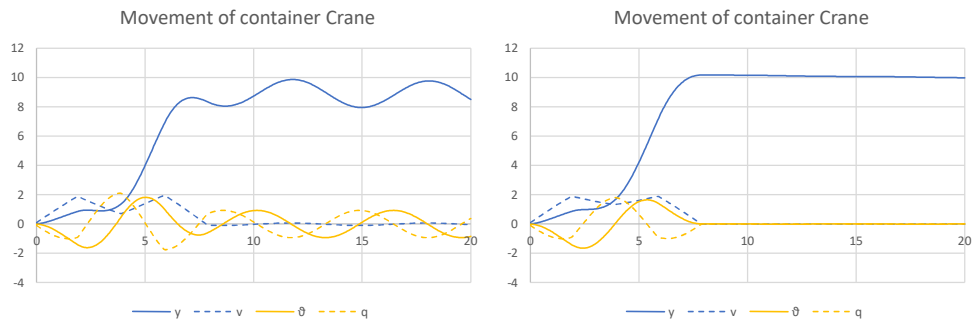

Figure 4. Movement of crane and container before and after optimisation

The solution corresponding to the 2sec timing for control inputs set the reduced loads to 17.5% of the main loads applied. The key finding of the work, though, was not the solution; it was that Excel now offers a practical method of solving such problems without needing to move to commercial dynamic simulation software.

**4.3  ModelOff 2013 Challenge Problem**

The second use case for modern Excel is a cash flow forecasting problem drawn from Round 2 of the 2013 ModelOff competition and made available by Diamuid Early [Early, 2023] with a recorded presentation of the solution he would build. What is shown here is the same problem refactored to use dynamic arrays and Lambda functions.

The original solution was presented as an assumptions sheet provided by the competition organisers, followed by the calculations as developed by Early. The forecast was presented as a number of blocks covering different aspects of the problem. The blocks included

- Timing
- Income statement
- Debt schedule
- Working capital – receivables
- Working capital – payables
- Cash Flow

A characteristic of the refactored solution is that rather than having formulas for individual cells or even line items, there are six main functions, each returning an entire block of the financial model. The function name is chosen to capture the intent of the block and the arguments provide a comprehensive list of references.

```
= TimingBlockλ(modelStart, modelDuration)
= IncomeStatementλ(unitVolume, unitPrice, unitDirectCosts,
      depreciationMonthly, interest)
= DebtScheduleλ(modelStart,    modelDuration,
      debtAmortisationDate,    debtFacilityA,
      debtAmortisationAmount, interestRate / 12)
= WorkingCapitalλ(revenue,     cashReceiptTiming)
= WorkingCapitalλ(-costs,      cashPaymentTiming)
= CashFlowλ(
      modelDuration,
      openingAccReceivable,    openingCashReceiptTiming,
            AccountsReceivable#,
      openingAccountsPayable, openingCashPaymentTiming,
            AccountsPayable#,
      interest, repayment, openingCash
  )
```

The first function **TimingBlockλ** requires a start date and the number of years the model is to run for. The elements of the timing block form a reasonably conventional pattern with a counter row and monthly periods characterized by opening and closing dates. Slightly more problem specific was the decision to include year and quarter values for each period.

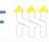

Figure 5. Timing Data for ModelOff Challenge Problem

```
TimingBlockλ
= LET(
    counter,     SEQUENCE(1, 12 * modelDuration),
    periodEnd,   EOMONTH(modelStart, counter - 1),
    periodStart, 1 + EOMONTH(+periodEnd, -1),
    year,        YEAR(periodEnd),
    quarter,     1 + QUOTIENT(MONTH(periodEnd) - 1, 3),
    VSTACK(counter, periodStart, periodEnd, year, quarter)
  )
```

By using **VSTACK** to output the counter along with period start and ends, as well as specific data such as year and quarter, the various detailed formulas are brought together in a way that is easily read and checked and the number of formula cells is minimised; the smaller the count of formula cells, the less is the opportunity for error.

It is also possible to generalise such formulas further in order to turn them into generic library functions by including steering data to select the output options and the unit of time represented by a period. In the present case, however, application-specific functionality was sufficient.

The next area of particular interest was the working capital block. The first point of note is that both the accounts receivable and accounts payable were summarised by the same Lambda function.

```
= LET(
    cashChange,  (-1) *TAKE(
                    Convolveλ(timing, amounts),
                    12 * modelDuration),
    closing,    SCAN(0, amounts + cashChange, Addλ),
    opening,    closing - (amounts + cashChange),
    VSTACK(opening, amounts, cashChange, closing)
)
```

The key to the formula is the imported Lambda function **Convolveλ** that generates a convolution of two arrays. Convolutions are used in satellite signal processing to support GPS, image processing and many other high-tech applications. To maintain efficiency for extremely large arrays of data the function is based upon Fast Fourier Transforms. The content is the preserve of professional mathematicians and electrical engineers and would normally be taught over the course of several specialist university lectures.

Despite the complexity of the implementation, the idea of a convolution is relatively simple when applied to small problems such as the present case. Basically, all that is required is to write the two arrays, one as a row and the other as a column and take the outer product of the two arrays. The result sought is the sum of terms arising on the reverse diagonals, as shown in the figure below.

|     | 612,296 | 612,296 | 612,296 | 363,879 | 363,879 | 363,879 | 272,909 | 272,909 | 272,909 | 545,818 | 545,818 | 545,818 |
|-----|---------|---------|---------|---------|---------|---------|---------|---------|---------|---------|---------|---------|
| -   | 0       | 0       | 0       | 0       | 0       | 0       | 0       | 0       | 0       | 0       | 0       | 0       |
| 60% | 367,378 | 367,378 | 367,378 | 218,327 | 218,327 | 218,327 | 163,745 | 163,745 | 163,745 | 327,491 | 327,491 | 327,491 |
| 25% | 153,074 | 153,074 | 153,074 | 90,970  | 90,970  | 90,970  | 68,227  | 68,227  | 68,227  | 136,455 | 136,455 | 136,455 |
| 15% | 91,844  | 91,844  | 91,844  | 54,582  | 54,582  | 54,582  | 40,936  | 40,936  | 40,936  | 81,873  | 81,873  | 81,873  |

|         | 0       | 0       | 0       | 0       | 0       |
|---------|---------|---------|---------|---------|---------|
|         | 367,378 | 218,327 | 218,327 | 218,327 | 163,745 |
|         | 153,074 | 153,074 | 90,970  | 90,970  | 90,970  |
|         | 91,844  | 91,844  | 91,844  | 54,582  | 54,582  |
|         | 612,296 | 463,246 | 401,141 | 363,879 | 309,297 |

**Figure 6. Calculating convolutions arising from payment schedules or depreciation.**

The point of the example is that it is perfectly possible to use imported Lambda functions as black boxes; there is no need to understand the details of the method provided the function has been thoroughly tested and sampling the result shows it to give the correct results.

## 5    ELEMENTS OF RISK

An objective of this paper is to demonstrate that, using Excel 365, it is now possible to create solutions to problems that have little or no resemblance to past spreadsheet solutions. The approaches are so different that it would not be reasonable to assume that the risks are in any way similar. Whereas traditional spreadsheet solutions are highly interactive processes based on the cell-matrix paradigm, modern dynamic array solutions can have far more in common with formal programming. The only thing that still distinguishes a spreadsheet solution is its use of a 2D grid in which multidimensional array calculations can be performed and presented to the reader.

With these changes comes another problem. It is likely that solutions such as those presented here will be beyond the capabilities of the majority of end-users; they will have neither the time nor the inclination to attempt programming tasks of this nature. It is likely that applications built by developers will coexist with *ad-hoc* workbooks developed along traditional lines.

For traditional spreadsheets the risk is most likely covered by "if you do what you have always done, you will get what you always have" or, in other words, "spreadsheets that have errors like dogs have fleas". The more formal functional programming approach may well reduce the error rate for spreadsheets but, at present, there is little to no experience so that remains pure speculation. Nevertheless, if we consider Ray Panko's 2010 taxonomy of spreadsheet errors, one may hazard reasonable guesses as to what aspects of error creation should improve and where the risk might actually increase.

If we consider a development model in which the intended end user specifies the problem they wish to address and a separate developer implements the solution, then either the end user has to have reasonable expectations in terms of spreadsheet domain knowledge, or the developer must have significant application domain knowledge. Any gaps present a risk.

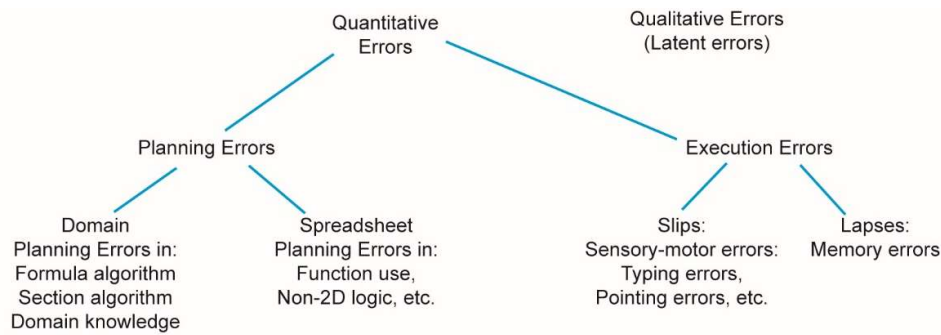

**Figure 6. Taxonomy of Spreadsheet Errors**

One thing that may help a modern Excel solution is the extent to which the use of named variables within the **LET** functions capture the intent of the solution rather than simply the actions of the author when creating the model. This should provide an element of built-in documentation to offset the fact that a modern solution can require the use of advanced programming concepts.

Another aspect of error reduction is testing. Excel, as yet, does not support key aspects of testing such as stepping through the code to check each line of code. Nevertheless, it should be possible to use unit testing to build confidence as one goes.

Something that does change, is the role of the end user. It is not realistic to expect them to perform cell by cell formula checks as they might have attempted in the past. The process was based upon the concept of code as a 'white box' and such checks served to improve confidence. Unfortunately, experience shows that such checks fail to find a majority of errors, so the additional confidence is largely unjustified. What the end user should do is focus on validation and use simple datasets constructed to demonstrate the code's fitness for purpose. The end user is the domain expert and it they who are better placed to determine the acceptability of any derived results. Conversely it is the developer that is responsible for line by line checks and unit testing as code verification activities.

As for execution errors, problems of typing and pointing should almost be eliminated. The array solutions do not allow errors here and there; either the result is correct everywhere or nothing is correct. In addition, reducing the number of formula cells from 6,000 to 10 (say) dramatically reduces the opportunity for random error.

In the author's experience, it is almost inevitable that a developer-built solution will exhibit 'carbuncles' after a period of use. As with any application, commercial or bespoke, the end user will from time to time think of additional requirements. It is likely that these will be implemented in the form of traditional formulas making direct reference to cells within the installed workbook. The changes will work and most likely give correct results as they are implemented, but they introduce the risk of latent error. The chances are that such extra code will not respond dynamically to new input data, so the code should be refactored to incorporate new requirements soon after they arise.

## 6     CONCLUSIONS

Modern Excel claims to be Turing complete and this paper has set out to demonstrate a range of techniques that a world away from the naive constructs, characteristic of traditional spreadsheet solutions. Although the techniques are far more advanced, they capture the intent of the programmer and allow solutions to be developed that would otherwise require VBA or Typescript. It is argued that semantically meaningful coding practices could give more reliable results.